\documentclass[aps,prd,twocolumn,nofootinbib]{revtex4-1}
\usepackage{epsfig}
\usepackage[colorlinks,linkcolor=blue,anchorcolor=blue,citecolor=blue,urlcolor=blue,breaklinks=true]{hyperref}
\usepackage{graphics}
\usepackage{slashed}
\usepackage{color}
\usepackage{amsmath}
\usepackage{bm}
\usepackage{setspace}

\begin{document}
\author{Ya-Peng Zhao$^{1}$}\email{zhaoyapeng2013@hotmail.com}
\author{Pei-Lin Yin$^{2}$}
\author{Zhen-Hua Yu$^{1}$}
\author{Hong-Shi Zong$^{1,3,4}$}\email{zonghs@nju.edu.cn}
\address{$^{1}$ Department of Physics, Nanjing University, Nanjing 210093, China}
\address{$^{2}$College of Science, Nanjing University of Posts and Telecommunications, Nanjing 210023, China}
\address{$^{3}$State Key Laboratory of Theoretical Physics, Institute of Theoretical Physics, CAS, Beijing, 100190, China}
\address{$^{4}$Joint Center for Particle, Nuclear Physics and Cosmology, Nanjing, 210093, China}
\title{Finite volume effects on chiral phase transition and pseudoscalar mesons properties from the Polyakov-Nambu-Jona-Lasinio model}

\begin{abstract}
 Within the framework of Polyakov-Nambu-Jona-Lasinio model and by means of Multiple Reflection Expansion, we study the finite volume effects on chiral phase transition, especially its influence on the location of the possible critical end point (CEP) and masses of mesons.
 Our result shows that as the radius of spherical volume decreases, the location of CEP shifts toward smaller temperature while changes little in chemical potential.
  As for the finite volume effects on the masses of mesons, the masses of $\pi$ and $K$ increase with decreasing volume, while for $\sigma$, $\eta$ and $\eta '$ the situation is the opposite.
  Especially, the masses of chiral parters $\pi$ and $\sigma$ get closer as the volume decreases, indicating that the dynamical chiral symmetry breaking effect reduces with decreasing volume.
\bigskip

\noindent Key-words: finite volume effects, chiral phase transition, pseudoscalar mesons, Polyakov-Nambu-Jona-Lasinio model
\bigskip

\noindent PACS Number(s): 12.38.Mh, 11.10.Wx, 64.60.an

\end{abstract}

\pacs{12.38.Mh, 12.39.-x, 25.75.Nq}

\maketitle

\section{INTRODUCTION}
Nowadays, it is still challenging to understand the hadron-quark phase transition, whether from the experimental or theoretical points of view. The fundamental theory for describing strong interactions is Quantum Chromodynamics (QCD). However, it is extremely difficult to solve from the first principle in the regime of intermediate temperature and chemical potential, e.g., lattice QCD~\cite{doi:10.1093/ptep/ptx018} confronts the sign problem. In this context, people resort to effective models, such as the Nambu-Jona-Lasinio (NJL)~\cite{RevModPhys.64.649,BUBALLA2005205,PhysRevD.71.116002} model. Its Lagrangian is constructed in such a way that the basic symmetries of QCD which are observed in nature are part and parcel of it. Especially it maintains the chiral symmetry, and at the same time shows how DCSB happens. But it also has a shortcoming being not able to describe quark confinement.

To construct a model by incorporating the quark confinement at low energies, the polyakov loop was introduced in the NJL model~\cite{FUKUSHIMA2004277} to simulate the quark confinement effects (the so-called PNJL model). In the PNJL model, quarks are not only coupled to the chiral condensate but also to the polyakov loop, so that we can study chiral phase transition and deconfinement phase transition at the same time.
This model has proven to be more successful in reproducing lattice data concerning QCD thermodynamics~\cite{PhysRevD.73.014019} than NJL model, because the coupling to the Polyakov loop produces a suppression of the unphysical colored quark states (one or two quark states) which should not contribute to the thermodynamics below the critical temperature. Following this model, many properties of the strongly interacting matter can be obtained, such as its phase diagram~\cite{1742-6596-312-1-012001,0034-4885-74-1-014001,PhysRevD.77.114028,PhysRevD.94.071503} and the properties of mesons~\cite{PhysRevD.75.065004,JPhysG.38.105003,PhysRevC.79.055208}.

 A comprehension of finite volume effects is very important for the analysis and interpretation of QCD simulations on a finite, discrete space-time lattice and ultra relativistic heavy ion collisions. Lattice QCD is a powerful method to study strong-interacting matter from the first principle of QCD. The
 development of calculation methods has made it possible to use pion and quark masses down to their physical values~\cite{DURR2011265,Colangelo2011,PhysRevLett.113.082001}. Thus, a thorough understanding of the volume effects becomes quite important because the smaller pion mass implies more important long-range effects, and therefore larger volume effect at the same lattice size. This is especially true close to the chiral phase transition, where the behavior of the system is dominated by the critical fluctuations of light degrees of freedom. The strong interacting matter produced by ultra relativistic heavy ion collisions is finite in volume, and its size depends on the nature of the colliding nuclei, the center of mass energy and the centrality of collision. The volume of homogeneity before freeze-out for Au-Au and Pb-Pb collisions ranges between approximately $50\sim250$ $\mathrm{fm}^{3}$~\cite{PhysRevC.85.044901} based on the UrQMD transport approach~\cite{BASS1998255}. And the smallest quark-gluon plasma (QGP) system produced at RHIC could be as low as $(2\ \mathrm{fm})^{3}$ as the Ref.~\cite{JPhysG.38.085101} estimated.

 In this paper, what we are particularly interested in is to analyze how the chiral phase transition and masses of mesons in a strongly interacting matter depend on the volume of the system. To incorporate finite volume effects different procedures have been employed, such as Dyson-Schwinger equations with the anti-periodic  boundary condition~\cite{Shi:2018swj}, the renormalization group approach~\cite{PhysRevD.73.074010,KLEIN20171}, NJL model with stationary wave solution~\cite{Wang:2018ovx}, and PNJL model with a low momentum cutoff $\Lambda$ on the thermodynamics potential~\cite{IntJModPhysA.32.1750067,PhysRevD.87.054009}. Here we follow Ref.~\cite{Grunfeld2018} to incorporate the finite volume effects by MRE formalism~\cite{BALIAN1970401} and extend it to finite chemical potential. Compared to other methods, MRE describes the sphere instead of cubic, which is closer to the fireball produced by the relativistic heavy ion collisions. Therefore, the surface and curvature effects of the sphere are properly considered.

This paper is organized as follows: In Sec. II, we give a brief introduction to the PNJL model at finite temperature and finite quark chemical potential. With the help of scalar susceptibility we study the chiral phase transition, especially the influences of the finite volume effect on the behaviour of the CEP. In Sec. III, We mainly focus on the masses of the pseudoscalar mesons and $\sigma$ meson in a finite volume. Finally, we will give a brief summary in Sec. IV.

\section{Chiral phase transition and finite volume effects within $2+1$ flavors PNJL model}\label{ok}
The Lagrangian of $2+1$ flavors of the PNJL model reads~\cite{Grunfeld2018}
\begin{eqnarray}
\mathcal{L}_{\rm SU(3)} &=& \bar{\Psi}(i{\not\!D}-\hat{m})\Psi +\frac{g_{S}}{2}\sum_{a=0}^{8}\,[(\bar{\Psi}\lambda^a\Psi)^2+(\bar{\Psi}i\gamma_5\lambda^a\Psi)^2] \nonumber \\
&&\,+g_{D}[\rm det\bar{\Psi}(1+\gamma_5)\Psi+\rm det\bar{\Psi}(1-\gamma_5)\Psi]\nonumber \\
&&\,-\mathcal{U}(\Phi,\bar{\Phi};T), \nonumber \\
\label{suthreelag}
\end{eqnarray}
where $\Psi=(u,d,s)$ represents the three flavor quark field with three colors and $\hat{m}=diag(m_{u},m_{d},m_{s})$ stands for the current quark mass matrix. Here, we assume the $SU(2)_{V}$ isospin symmetry which means $m_{u}=m_{d}$. $g_{S}$ is the effective coupling strength of four point interaction of quark fields and $g_{D}$ is the six-quark interaction coupling, which breaks the axial $U_{A}(1)$ symmetry. The normalized color-traced Polyakov loop expectation value and its Hermitian conjugation defined as
\begin{eqnarray}
\Phi=\frac{\langle Tr_{c}L \rangle}{N_{c}},     \bar{\Phi}=\frac{\langle Tr_{c}L^{\dagger} \rangle}{N_{c}},
\end{eqnarray}
The Polyakov line is represented as
\begin{eqnarray}
L(\vec{x})=Pexp(i\int_{0}^{\beta}A_{4}(\vec{x},\tau)d\tau),
\end{eqnarray}
where $A_{4}=iA_{0}$ is the temporal component of Euclidian gauge field $(\vec{A},A_{4})$, $\beta=\frac{1}{T}$, and $P$ denotes the path ordering.
The covariant derivative is determined as
\begin{eqnarray}
D_{\mu}&=&\partial_{\mu}-iA_{\mu},\nonumber \\
A_{\mu}&=&\delta_{\mu}^{0}A_{0},
\end{eqnarray}
Here $A_{\mu}=gA_{\mu}^{a}\frac{\lambda^{a}}{2}$ and $g$ is the $SU(3)_{c}$ gauge coupling. The $\lambda^{a}$ stand for the Gell-Mann matrices with $\lambda^{0}=\sqrt{\frac{2}{3}}1$. The effective Polyakov potential $\mathcal{U}(\Phi,\bar{\Phi};T)$ that accounts for gauge field self-interactions we used in this work is
\begin{eqnarray}
\frac{\mathcal{U}(\Phi,\bar{\Phi};T)}{T^{4}}=-\frac{b_{2}(T)}{2}\bar{\Phi}\Phi-\frac{b_{3}}{6}(\Phi^{3}+\bar{\Phi}^{3})+\frac{b_{4}}{4}(\bar{\Phi}\Phi)^{2},
\end{eqnarray}
The expansion coefficients are determined by fitting several thermodynamics quantities as functions of temperature obtained in lattice QCD. A temperature-dependent coefficients
\begin{eqnarray}
b_{2}(T)=a_{0}+a_{1}(\frac{T_{0}}{T})+a_{1}(\frac{T_{0}}{T})+a_{3}(\frac{T_{0}}{T})^{3},
\end{eqnarray}
The corresponding parameters we used here from~\cite{Grunfeld2018} are given in Table \ref{tb1}.
Actually, one can expect $T_{0}=270$ $\mathrm{MeV}$ just in a pure gauge sector. For $2+1$ flavors with a current strange quark mass $m_{s}\approx150$ $\mathrm{MeV}$, this temperature is rescaled to about 187 $\mathrm{MeV}$, with an uncertainty about 30 $\mathrm{MeV}$ as Ref.~\cite{PhysRevD.76.074023} shows. In this work, we will take $T_{0}=270$ $\mathrm{MeV}$ and $T_{0}=185$ $\mathrm{MeV}$ to test the impact of $T_{0}$ on our results.
\begin{center}
\begin{table}
\caption{Parameter set used in our work.}\label{tb1}
\begin{tabular}{p{1.1cm} p{1.1cm} p{1.1cm} p{1.1cm} p{1.1cm} p{0.8cm} p{1.2cm}}
\hline\hline
$a_0$&$a_1$&$a_2$&$a_3$&$b_3$&$b_4$&$T_0(\mathrm{MeV})$\\
\hline
6.76&-1.95&2.625&-7.44&0.75&7.5&185\\
\hline\hline
\end{tabular}
\end{table}
\end{center}

 The NJL model parameters fixed at $T=\mu=0$ are not affected by introduction of the polyakov loop coupling, because the polyakov loop coupling appears only in the thermal part. The widely accepted parameters set according to Hatsuda and Kunihiro ~\cite{HATSUDA1994221} are given in Table \ref{tb2}.
\begin{center}
\begin{table}
\caption{Parameter set used in our work.}\label{tb2}
\begin{tabular}{p{1.7cm} p{1.7cm} p{1.7cm} p{1.3cm} p{1.3cm}}
\hline\hline
$\Lambda(\mathrm{MeV})$&$m_{ud}(\mathrm{MeV})$&$m_{s}(\mathrm{MeV})$&$g_{S}\cdot\Lambda^{2}$&$g_{D}\cdot\Lambda^{5}$\\
\hline
631.4&5.5&135.7&3.67&-9.29\\
\hline\hline
\end{tabular}
\end{table}
\end{center}
  Under the mean-field approximation, the thermodynamic potential density function is~\cite{PhysRevD.77.054023}
\begin{eqnarray}
\Omega(\mu,T,M_{f},\Phi,\bar{\Phi})&=&\mathcal{U}(\Phi,\bar{\Phi};T)+g_{S}\sum_{f=u,d,s}\sigma_{f}^{2}+4g_{D}\sigma_{u}\sigma_{d}\sigma_{s}\nonumber\\
&-&T\sum_{n}\int\frac{{\rm d}^3p}{(2\pi)^3}{\rm Tr}ln\frac{S^{-1}(i\omega_{n},\vec{p})}{T},
\end{eqnarray}
where $\sigma_{f}=\langle\bar{\Psi}_{f}\Psi_{f}\rangle$ denotes chiral condensate of the quark with flavor $f$. It relates to the constitute quark mass $M_{f}$ as
\begin{eqnarray}
M_{u}=m_{u}-2g_{S}\sigma_{u}-2g_{D}\sigma_{u}\sigma_{s},
\end{eqnarray}
\begin{eqnarray}
M_{s}=m_{s}-2g_{S}\sigma_{s}-2g_{D}\sigma_{u}\sigma_{u},
\end{eqnarray}
$\omega_{n}=\pi T(2n+1)$ are the Matsubara frequencies of fermions, and $S^{-1}$ is the inverse quark propagator
\begin{eqnarray}
S^{-1}(p_{0},\vec{p})=\gamma_{0}(p^{0}+\mu-iA_{4})-\vec{\gamma}\cdot\vec{p}-M,
\end{eqnarray}
using the identity ${\rm Tr}$ $ln(X)=ln$ $det(X)$ , we get
\begin{eqnarray}
\Omega(\mu,T,M_{f},\Phi,\bar{\Phi})&=&\mathcal{U}(\Phi,\bar{\Phi};T)+g_{S}\sum_{f=u,d,s}\sigma_{f}^{2}\\
&+&4g_{D}\sigma_{u}\sigma_{d}\sigma_{s}-6\sum_{f}\int_{0}^{\Lambda}\frac{{\rm d}^3p}{(2\pi)^3}E_{pf}\nonumber\\
&-&2T\sum_{f}\int_{0}^{\infty}\frac{{\rm d}^3p}{(2\pi)^3}(lnF^{+}_{f}+lnF^{-}_{f}),\nonumber
\end{eqnarray}
with
\begin{eqnarray}
F^{+}_{f}&=&1+3(\Phi+\bar{\Phi}e^{-\frac{E_{pf}-\mu}{T}})e^{-\frac{E_{pf}-\mu}{T}}+e^{-3\frac{E_{pf}-\mu}{T}},\nonumber\\
F^{-}_{f}&=&1+3(\bar{\Phi}+\Phi e^{-\frac{E_{pf}+\mu}{T}})e^{-\frac{E_{pf}+\mu}{T}}+e^{-3\frac{E_{pf}+\mu}{T}},
\end{eqnarray}
and $E_{pf}=\sqrt{p^{2}+M_{f}^{2}}$ is the single quasiparticle energy. In the above integrals, as Ref.~\cite{PhysRevC.79.055208}, the vacuum integral has a cutoff $\Lambda$ whereas the medium dependent integrals have been extended to infinity.

For any given $\mu$ and $T$, the behavior of the order parameters is obtained by minimizing the thermodynamic potential function $\Omega$ directly.
Now we are ready to introduce the effects of finite volume in the thermodynamic potential by means of the MRE formalism~\cite{PhysRevD.67.085010,PhysRevD.72.054009,PhysRevD.50.3328}. In the case of a finite spherical droplet it modifies the density of states as follows
\begin{eqnarray}
\rho_{i,MRE}(p,m_{i},R)=1+\frac{6\pi^{2}}{pR}f_{i,S}+\frac{12\pi^{2}}{(pR)^{2}}f_{i,C},
\end{eqnarray}
where $f_{i,S}$ denote the surface contribution to the density of states
\begin{eqnarray}
f_{i,S}=-\frac{1}{8\pi}(1-\frac{2}{\pi}\mathrm{arctan}\frac{p}{m_{i}}),
\end{eqnarray}
and the curvature contribution is given by Madsen$^{,}$s ansatz~\cite{PhysRevD.50.3328}
\begin{eqnarray}
f_{i,C}=\frac{1}{12\pi^{2}}[1-\frac{3p}{2m_{i}}(\frac{\pi}{2}-\mathrm{arctan}\frac{p}{m_{i}})],
\end{eqnarray}
which takes the finite quark mass contribution into account.

For massive quarks, the MRE density of states would become negative for a range of small momentum. Consequently, in order to obtain the thermodynamic quantities in a finite volume we remove this non-physical negative values by introducing an infrared (IR) cutoff in momentum space~\cite{PhysRevD.72.054009,PhysRevC.88.045803}. The following replacement must be performed.
\begin{eqnarray}
\int_{0}^{\Lambda,\infty}\frac{ d^{3}p}{(2\pi)^3}\cdots\rightarrow\int_{\Lambda_{i,IR}}^{\Lambda,\infty}\frac{ d^{3}p}{(2\pi)^3}\rho_{i,MRE}\cdots,
\end{eqnarray}
where the $IR$ cutoff $\Lambda_{i,IR}$ is the largest solution of the equation $\rho_{i,MRE}(p,m_{i},R)=0$ with respect to the momentum $p$.
\begin{figure}
\includegraphics[width=0.47\textwidth]{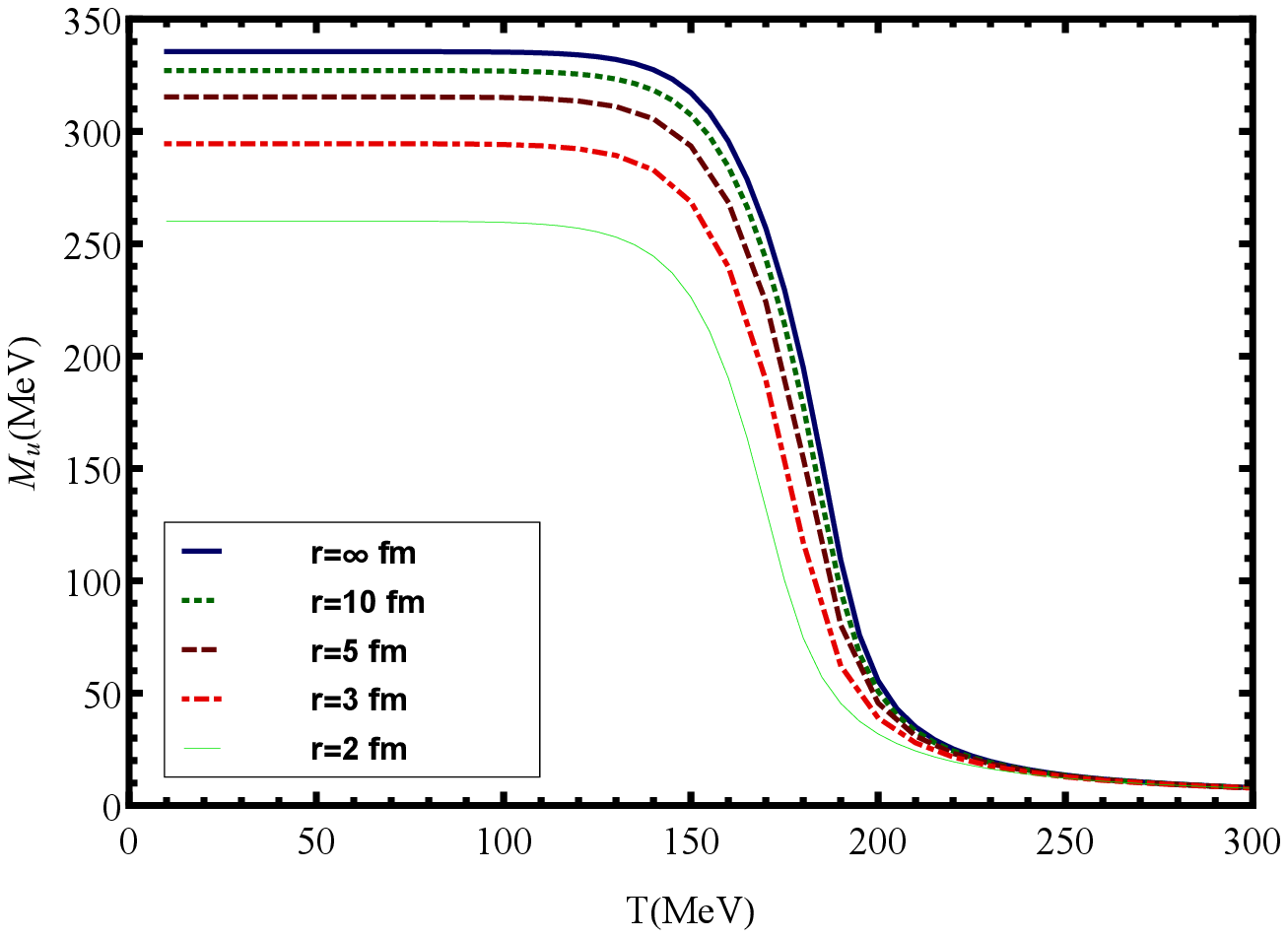}
\caption{Constituent mass $M_{u}$ as a function of $T$ at $\mu=0$ for different radius $r$.}
\label{Fig:1}
\end{figure}
\begin{figure}[h]
\includegraphics[width=0.47\textwidth]{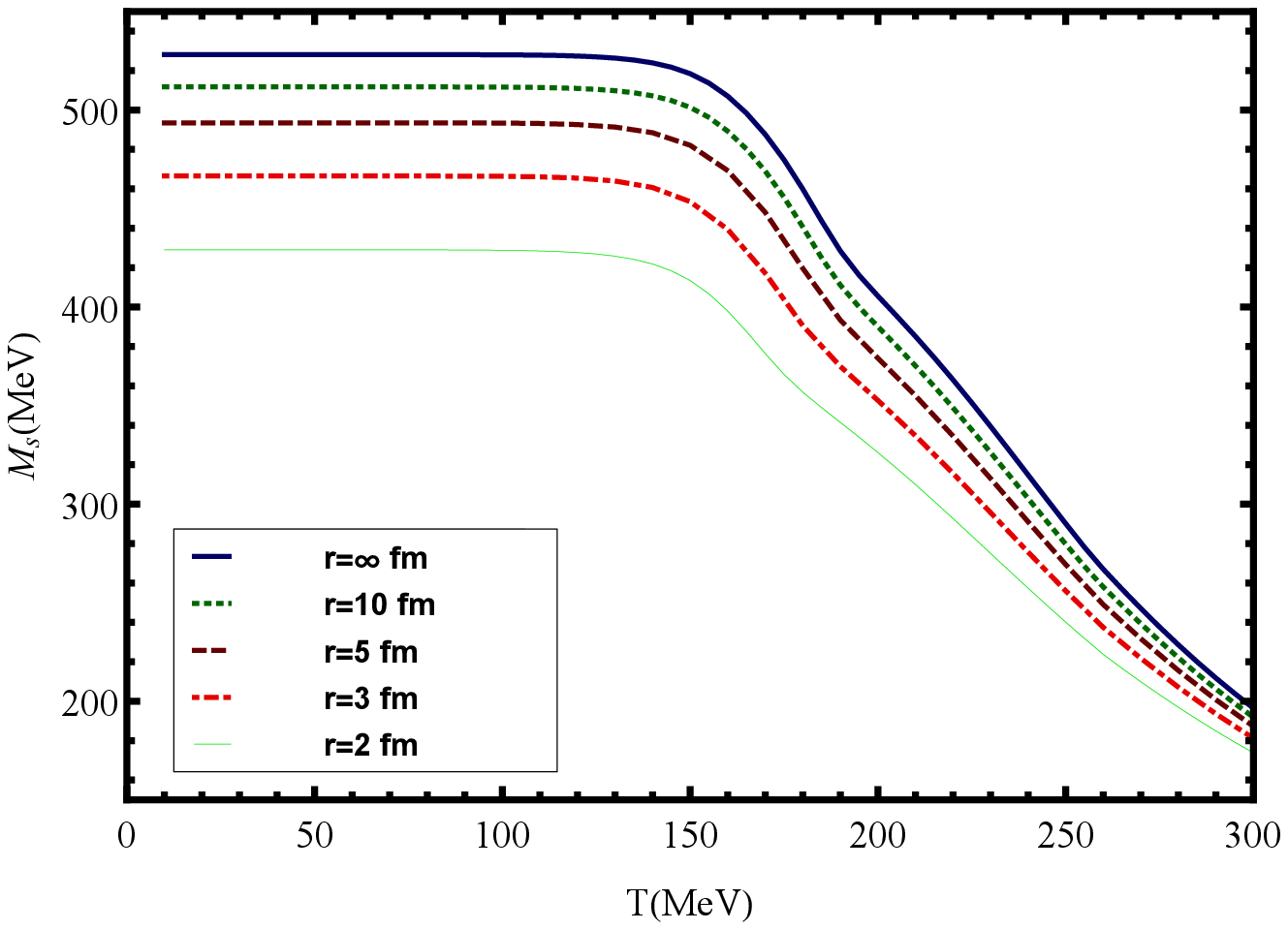}
\caption{Constituent mass $M_{s}$ as a function of $T$ at $\mu=0$ for different radius $r$.}
\label{Fig:2}
\end{figure}
\begin{figure}[h]
\includegraphics[width=0.47\textwidth]{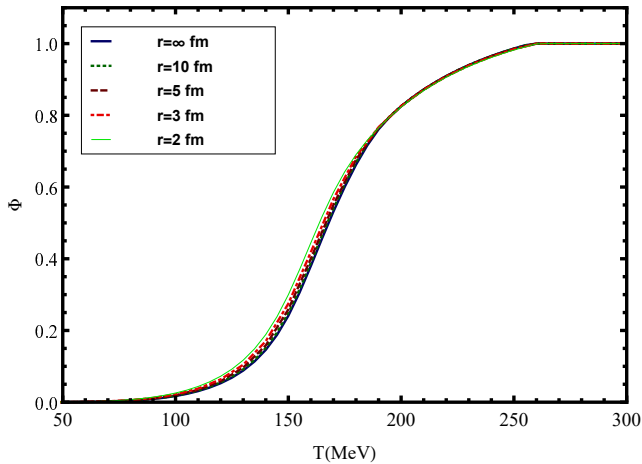}
\caption{Polyakov loop $\Phi$ as a function of $T$ at $\mu=0$ for different radius $r$.}
\label{Fig:3}
\end{figure}
\begin{figure}[h]
\includegraphics[width=0.47\textwidth]{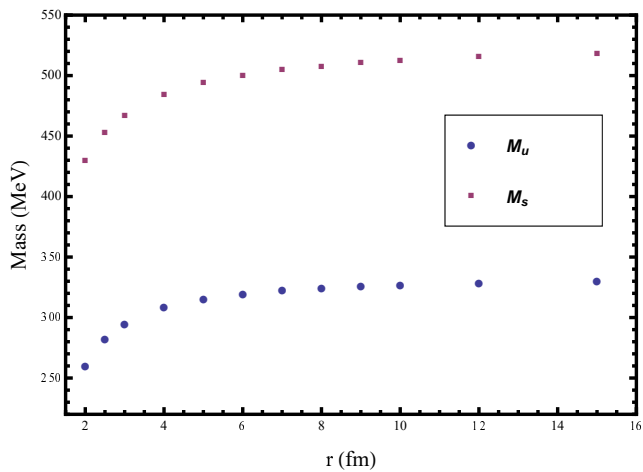}
\caption{Constituent masses $M_{u}$, $M_{s}$  as a function of radius $r$ at $T=0$ and $\mu=0$.}
\label{Fig:4}
\end{figure}

Now, through our numerical results, In Fig. \ref{Fig:1} and Fig. \ref{Fig:2} we plot the temperature and finite volume dependence of the constituent quark masses. Firs of all, we find the masses of constituent quarks $u$ and $s$ show strong volume dependence. In the low temperature region, when the volume decreases from infinity to a radius of 2 $\mathrm{fm}$, the mass of $u$ quark drops from 336 $\mathrm{MeV}$ to 260 $\mathrm{MeV}$ and $s$ quark drops from 528 $\mathrm{MeV}$ to 429 $\mathrm{MeV}$. This indicate that the DCSD effects reduce with decreasing volumes. Moreover, if the radius of spherical volume goes larger than 10 $\mathrm{fm}$, the volume effect can be ignored safely. Note that, strictly speaking, DCSB occurs for idealized systems that are infinitely large. However, in a finite volume V, if the condition
\begin{eqnarray}
V\cdot m\cdot\langle\bar{\Psi}\Psi\rangle\gg \pi ,
\end{eqnarray}
is satisfied, one can still observe the spontaneous formation of quark condensate and the dynamical mass generation~\cite{PhysRevD.73.034029}. In Fig. \ref{Fig:3} we plot the dependence of the Polyakov loop $\Phi$ on temperature and finite volume, in contrast to quarks, it shows a weak volume dependence. This result is similar to that obtained with PNJL model by a Monte-Carlo approach~\cite{PhysRevD.81.114017} and Ref.~\cite{PhysRevD.87.054009}. In Fig. \ref{Fig:4}, as a supplement to Fig. \ref{Fig:1} and \ref{Fig:2}, we show constitute quark mass varies with volume at zero temperature and quark chemical potential. It shows that the constitute quark mass decreases as the volume decreases, and this is exactly the opposite of its tendency to change with temperature. So, to some extent, reducing the volume and increasing the temperature have a similar effect on the chiral phase transition.

Moreover, the strongly interacting matter is expected to undergo a phase transition from hadronic phase to QGP phase at high temperatures and densities. For infinite volume, a popular scenario favors the phase transition is of the first order at sufficiently high chemical potential and one can observe a gap in the order parameter.
 At some smaller $\mu$, there exist a critical end point (CEP) where the first order phase transition ends and the system undergoes a second order transition. At even smaller $\mu$ we only have a crossover. The search for the position or even the existence of such a CEP is extremely important for theoretical and experimental physics because it marks a firm milestone in our understanding of the QCD phase diagram. Here, we study the volume effects on the location of the CEP.

\begin{figure}
\includegraphics[width=0.47\textwidth]{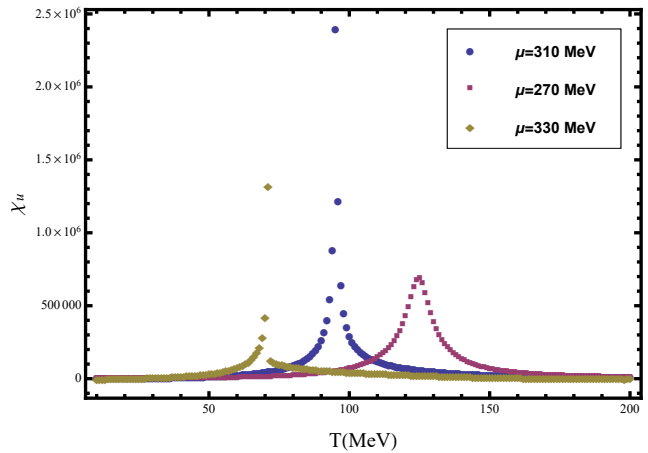}
\caption{The temperature susceptibility $\chi_{u}$ as a function of $T$ for three different $\mu$ in an infinity volume}
\label{Fig:5}
\end{figure}

\begin{figure}
\includegraphics[width=0.47\textwidth]{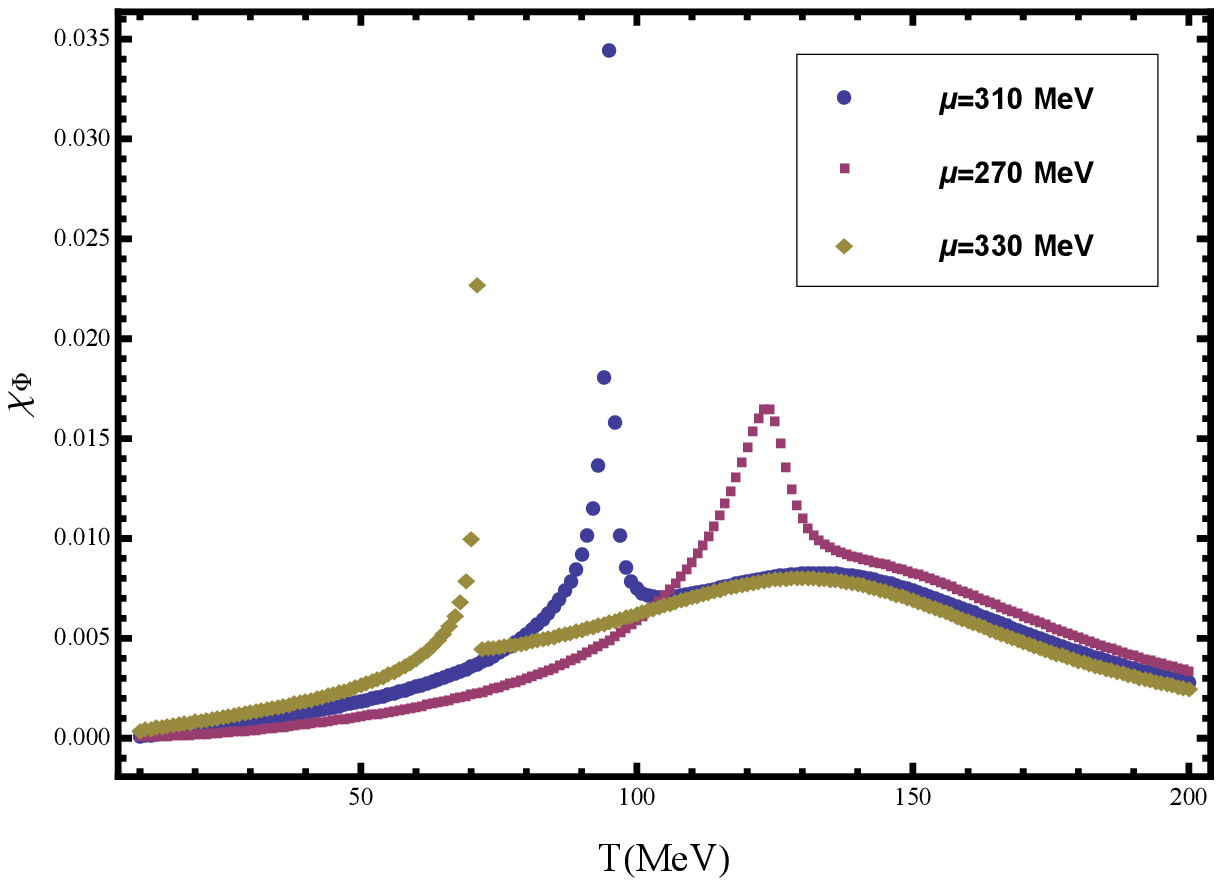}
\caption{The temperature susceptibility $\chi_{\Phi}$ as a function of $T$ for three different $\mu$ in an infinity volume}
\label{Fig:6}
\end{figure}

In order to determine the location of the CEP, we introduce susceptibility of condensate $\sigma_{u}$ and Polyakov loop $\Phi$ to the linear responses of temperature defined as Ref.~\cite{PhysRevD.76.074023}
\begin{eqnarray}\label{propagator}
\chi_{u}=\frac{\partial\sigma_{u}}{\partial T},\ \ \ \ \ \ \ \chi_{\Phi}=\frac{\partial\Phi}{\partial T},
\end{eqnarray}
Their behavior in an infinite volume are ploted in Fig. \ref{Fig:5} and Fig. \ref{Fig:6}. We can see that a quite sharp and divergent peak which corresponds to the CEP. Note that the CEP of deconfinement and chiral phase transition are in the same position.

\begin{figure}
\includegraphics[width=0.47\textwidth]{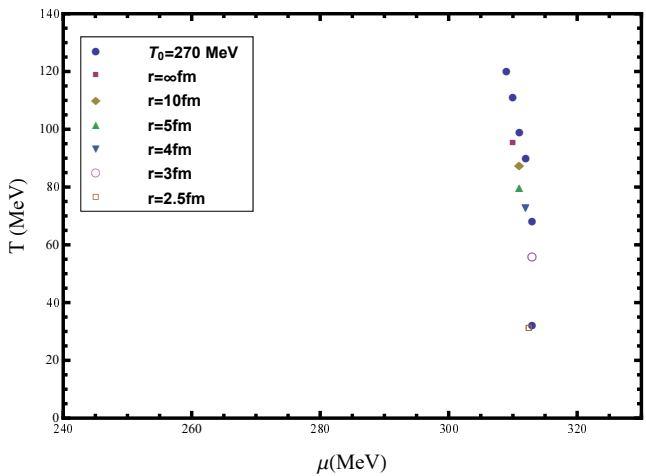}
\caption{Trajectory of volume dependent CEP, the blue circles represent our result obtained at $T_{0}=270\ \mathrm{MeV}$, the colours represent $T_{0}=185\ \mathrm{MeV}$}
\label{Fig:7}
\end{figure}

In a finite volume, however, the singularities are smoothed out and the susceptibilities have a finite peak near the infinite volume transition point~\cite{PhysRevLett.68.1738}. So, strictly speaking, we have a pseudo-CEP in a finite volume. Here, we choose the end point of first order phase transition as CEP like the case of infinite volume. According to this definition, we find that the coincidence of these two CEPs is independent of the volumes.
From Fig. \ref{Fig:7} we display the dependence of the CEP on volumes with different parameters, $T_{0}=270\ \mathrm{MeV}$ and $T_{0}=185\ \mathrm{Mev}$. Firstly, it clearly shows that the direction of the CEP shift with volume does not depend on our choice of $T_{0}$. Secondly, we note that the CEP shifts toward smaller temperatures but stay almost constant quark chemical potential, $\mu\simeq312\ \mathrm{Mev}$, when the radius of spherical volume decreases. This is meaningful for the CEP search in the heavy-ion experiments because, according to our calculation, if the radius of QGP is as small as 3 $\mathrm{fm}$, the $T$ of the CEP dropped from $95\ \mathrm{MeV}$ to $56\ \mathrm{MeV}$. Then we should try to find the CEP at low $T$ region. There also exists some possibility that if the radius gets smaller than about 2 $\mathrm{fm}$, we can not see any discontinuous of susceptibilities at all. This means there is no CEP, and the whole phase diagram indeed becomes a crossover. This behavior of the CEP is a little different from the Quark-Meson model~\cite{PhysRevD.90.054012} and Dyson-Schwinger equations~\cite{LI2019298,Shi:2018tsq}. In those studies, the $\mu$ of CEP shifts rapidly toward higher values with decreasing volume as well. On the other hand, our result is in qualitatively agreement with~\cite{PhysRevD.87.054009}

\section{Meson masses in a finite volume}\label{bigtwo}
 It is important to study the properties of mesons, propagating in a hot or dense medium. Because the degeneracy of the respective chiral partners~\cite{PhysRevD.70.116013,PhysRevD.79.116003,Blanquier:2015ina}, especially the lightest partners of $\pi$ and $\sigma$, can indicate an effective restoration of chiral symmetry. So, in this section, we mainly focus on the volume effects on the masses of pseudoscalar mesons and $\sigma$, at finite temperature and chemical potential. A detailed account of the calculational procedure for meson masses can be found in Ref.~\cite{PhysRevD.71.116002}. Here we give the basic formula that we use.
For the pseudoscalar mesons, the pole mass can be determined by this condition
\begin{equation}
1-P_{M}\Pi_{M}^{P}(P_{0}=m_{M},\vec{P}=0)=0,
\end{equation}
Here $P_{M}$ is the effective coupling constants and $\Pi_{M}^{P}$ is the one-loop polarization function. For mesons $\pi$ and $K$ (actually, here we means $K^{+}$), the effective coupling constants as follows
\begin{eqnarray}
P_{\pi}=g_{S}+g_{D}\sigma_{s},\nonumber \\
P_{K}=g_{S}+g_{D}\sigma_{u},
\end{eqnarray}
and the polarization function of $\pi$ and $K$ are
\begin{eqnarray}
\Pi_{\pi}^{P}=2\Pi^{P}_{uu},\ \ \ \ \ \  \Pi_{K}^{P}=2\Pi^{P}_{su},
\end{eqnarray}
Where $\Pi^{P}_{ij}$ has the form
\begin{eqnarray}
\Pi^{P}_{ij}(P)&=&-iN_{c}\int\frac{d^{4}p}{(2\pi)^{4}}tr_{D}[S^{i}(p)\gamma_{5}S^{j}(p+P)\gamma_{5}]\\ &=&2iN_{c}(I_{1}^{i}+I_{1}^{j})-2iN_{c}[P^{2}-(M_{i}-M_{j})^{2}]I_{2}^{ij},\nonumber
\end{eqnarray}
$tr_{D}$ is the trace over Dirac matrices, and
\begin{eqnarray}
I^{i}_{1}=\int^{\Lambda}\frac{d^{4}p}{(2\pi)^{4}}\frac{1}{p^{2}-M_{i}^{2}},
\end{eqnarray}
\begin{eqnarray}
I^{ij}_{2}=\int^{\Lambda}\frac{d^{4}p}{(2\pi)^{4}}\frac{1}{[p^{2}-M_{i}^{2}][(p+P)^{2}-M_{j}^{2}]},
\end{eqnarray}
Next, in order to determine the pole mass of $\eta$ and $\eta '$, we should solve a matrix equation
\begin{eqnarray}
det|1-P \Pi^{P}|=0,
\end{eqnarray}
with
\begin{equation}
\Pi^{P}=\left[
\begin{array}{cc}
\frac{2}{3}[2\Pi^{uu}+\Pi^{ss}]&\frac{2\sqrt{2}}{3}[\Pi^{uu}-\Pi^{ss}]\\
\frac{2\sqrt{2}}{3}[\Pi^{uu}-\Pi^{ss}]&\frac{2}{3}[\Pi^{uu}+2\Pi^{ss}]\\
\end{array}
\right],
\end{equation}
and
\begin{equation}
P=\left[
\begin{array}{cc}
g_{S}-\frac{2}{3}g_{D}(2\phi_{u}+\phi_{s})&\frac{\sqrt{2}}{3}g_{D}(\phi_{u}-\phi_{s})\\
\frac{\sqrt{2}}{3}g_{D}(\phi_{u}-\phi_{s})&g_{S}+\frac{1}{3}g_{D}(4\phi_{u}-\phi_{s})\\
\end{array}
\right],
\end{equation}
The matrix equation can be translated into
\begin{equation}\label{constrains}
  \left\{\begin{array}{lcl}
           M_{\eta}^{-1}(m_{\eta},\vec{0})=\mathcal{A}+\mathcal{C}-\sqrt{(\mathcal{A}-\mathcal{C})^{2}+4\mathcal{B}^{2}}=0,\\
           M_{\eta'}^{-1}(m_{\eta'},\vec{0})=\mathcal{A}+\mathcal{C}+\sqrt{(\mathcal{A}-\mathcal{C})^{2}+4\mathcal{B}^{2}}=0,
         \end{array}\right.
\end{equation}
with $\mathcal{A}=P_{88}-\Delta\Pi_{00},\mathcal{C}=P_{00}-\Delta\Pi_{88},\mathcal{B}=-(P_{08}+\Delta\Pi_{08})$, and $\Delta=P_{00}P_{88}-P_{08}^{2}$.

For $\sigma$, the procedure is exactly the same with $\eta$ and $\eta '$. What we need is to replace the pseudoscalar polarization functions by the scalar ones and the effective coupling constant $P$ by $S$. That is to say
\begin{eqnarray}
\Pi^{S}_{ij}(P)&=&iN_{c}\int\frac{d^{4}p}{(2\pi)^{4}}tr_{D}[S^{i}(p)S^{j}(p+P)]\\ &=&2iN_{c}(I_{1}^{i}+I_{1}^{j})-2iN_{c}[P^{2}-(M_{i}+M_{j})^{2}]I_{2}^{ij},\nonumber
\end{eqnarray}
and
\begin{equation}
S=\left[
\begin{array}{cc}
g_{S}+\frac{2}{3}g_{D}(2\phi_{u}+\phi_{s})&-\frac{\sqrt{2}}{3}g_{D}(\phi_{u}-\phi_{s})\\
-\frac{\sqrt{2}}{3}g_{D}(\phi_{u}-\phi_{s})&g_{S}-\frac{1}{3}g_{D}(4\phi_{u}-\phi_{s})\\
\end{array}
\right],
\end{equation}

From now on, we discuss meson properties at finite $T$ and $\mu$. The medium version of integral $I_{1}$ is defined as
\begin{equation}
I_{1}(T,\mu)=iT\sum_{j\in Z}\int\frac{d^{3}p}{(2\pi)^{3}}\frac{1}{(i\omega_{j}+\mu)^{2}-E^{2}_{p}},
\end{equation}
with the help of residue theorem
\begin{eqnarray}
I_{1}(T,\mu)&=&i\int\frac{d^{3}p}{(2\pi)^{3}}\sum_{i=1}^{2}Res_{z_{i}}(\frac{f(z)}{(z+\mu)^{2}-E_{p}^{2}})\\ &=&-i\int\frac{d^{3}p}{(2\pi)^{3}}\frac{1}{2E_{p}}(1-f(E_{p}-\mu)-f(E_{p}+\mu)),\nonumber
\end{eqnarray}
$f(E_{p}-\mu)$ and $f(E_{p}+\mu)$ are Fermi occupation numbers of quarks and antiquarks respectively
\begin{eqnarray}
f(E_{p}-\mu)=\frac{1}{e^{(E_{p}-\mu)/T}+1},\nonumber \\
f(E_{p}+\mu)=\frac{1}{e^{(E_{p}+\mu)/T}+1},
\end{eqnarray}
Similar to $I_{1}$, the medium version of integral $I_{2}$ is
\begin{eqnarray}
I_{2}^{ij}(P_{0},\vec{0})&=&i\int\frac{d^{3}p}{(2\pi)^{3}}\sum_{l=1}^{4}Res_{z_{l}}\nonumber \\
&&\frac{f(z)}{[(z+\mu)^{2}-E_{p_{i}}^{2}][(z+\mu+P_{0})^{2}-E_{p_{j}}^{2}]},
\end{eqnarray}
The solutions of Eqs. (19) and (25) are real values, as the obtained mass, when we consider a meson that is stable. At the opposite, when a meson is unstable, the polarization function is a complex function~\cite{PhysRevD.75.065004} and the mass becomes also a complex number, written as
\begin{equation}
m=m_{Re}-\frac{i}{2}\cdot\Gamma,
\end{equation}
$m_{Re}$ is the real part and is identified to the particle mass.

There is one point that we should pay attention to is, for consistence, only the vacuum part needs to be regularised. For the medium part, they are not divergent at all because of the occupation number density are given in terms of the Fermi distribution function. Always remember that the finite volume effects are introduced by Eq. (16).
According to Ref.~\cite{PhysRevD.75.065004}, the changes in going from NJL to PNJL model can then be summarized in the following prescriptions
\begin{eqnarray}
f(E_{p}-\mu)&\Rightarrow&f_{\Phi}^{+}(E_{p})\\
&=&\frac{(\bar{\Phi}+2\Phi e^{-\frac{E_{p}-\mu}{T}})e^{-\frac{E_{p}-\mu}{T}}+e^{-3\frac{E_{p}-\mu}{T}}}{1+3(\bar{\Phi}+\Phi e^{-\frac{E_{p}-\mu}{T}})e^{-\frac{E_{p}-\mu}{T}}+e^{-3\frac{E_{p}-\mu}{T}}},\nonumber
\end{eqnarray}
\begin{eqnarray}
f(E_{p}+\mu)&\Rightarrow&f_{\Phi}^{-}(E_{p})\\
&=&\frac{(\Phi+2\bar{\Phi} e^{-\frac{E_{p}+\mu}{T}})e^{-\frac{E_{p}+\mu}{T}}+e^{-3\frac{E_{p}+\mu}{T}}}{1+3(\Phi+\bar{\Phi} e^{-\frac{E_{p}+\mu}{T}})e^{-\frac{E_{p}+\mu}{T}}+e^{-3\frac{E_{p}+\mu}{T}}},\nonumber
\end{eqnarray}
\begin{figure}
\includegraphics[width=0.47\textwidth]{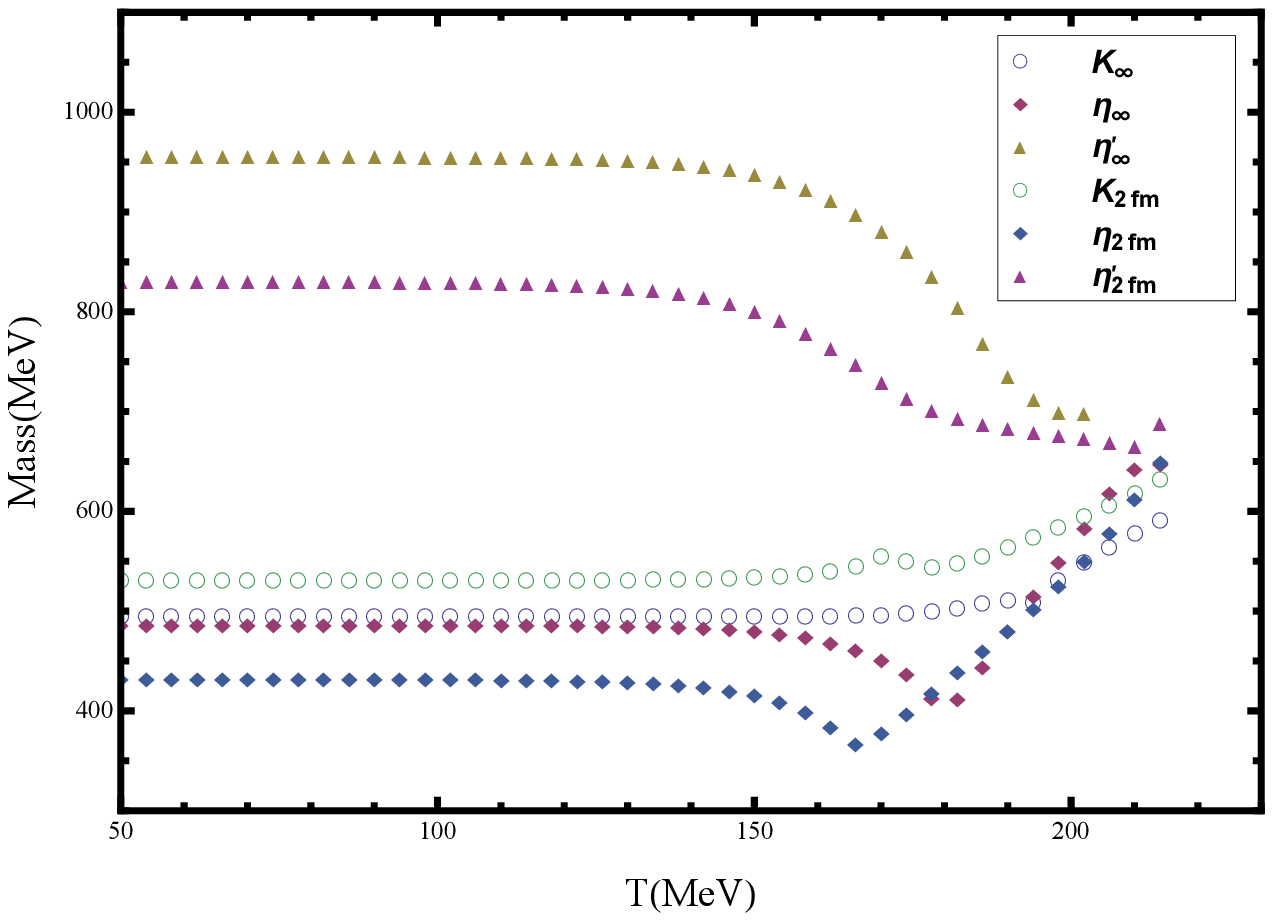}
\caption{The masses of $K$, $\eta$ and $\eta '$ as a function of $T$ at $\mu=0$ for two different volumes.}
\label{Fig:8}
\end{figure}
\begin{figure}
\includegraphics[width=0.47\textwidth]{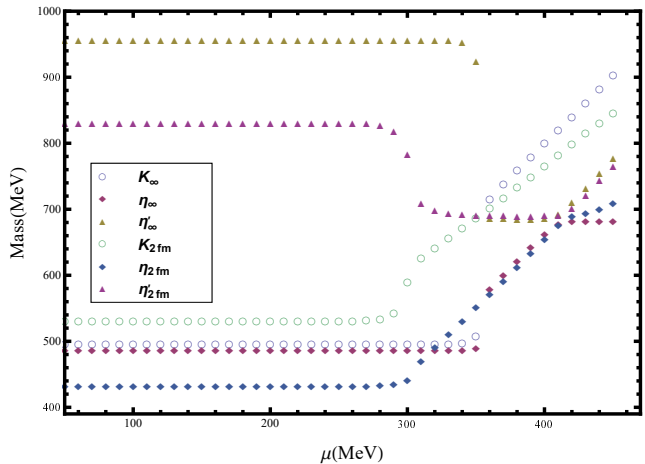}
\caption{The masses of $K$, $\eta$ and $\eta '$ as a function of $\mu$ at $T=0$ for two different volumes.}
\label{Fig:9}
\end{figure}

In the Fig. \ref{Fig:8} and Fig. \ref{Fig:9}, for convenience, we just plot the variation of meson masses with temperature and chemical potential in an infinite volume and a spherical volume with a radius of $2\ \mathrm{fm}$. Firstly, in an infinity volume, our results are in qualitatively agreement with~\cite{PhysRevD.79.116003}. The masses of mesons change continuously from the chiral symmetry broken phase to the chiral symmetry restored phase with temperature $T$, which means only a crossover happens. But on the $\mu$-axis, there appears a sudden discontinuity at a certain critical $\mu$, which indicates the first order phase transition happens. Secondly, in a spherical volume with a radius smaller than about $2\ \mathrm{fm}$, all the masses change continuously even on the $\mu$-axis. This, yet again, shows that a crossover happens in the whole phase diagram. To some extent, the masses of mesons reflecting an analogous behaviour of the chiral condensate $\langle\bar{\Psi}_{f}\Psi_{f}\rangle$.

In addition, at low temperature, we also find the mass of $K$ increases with higher temperatures and smaller volumes, but for $\eta$ and $\eta '$, the situation is just the opposite. This again shows that the decrease of volume restores the spontaneous breaking of chiral symmetry in a similar way as increase in temperature.

\begin{figure}
\includegraphics[width=0.47\textwidth]{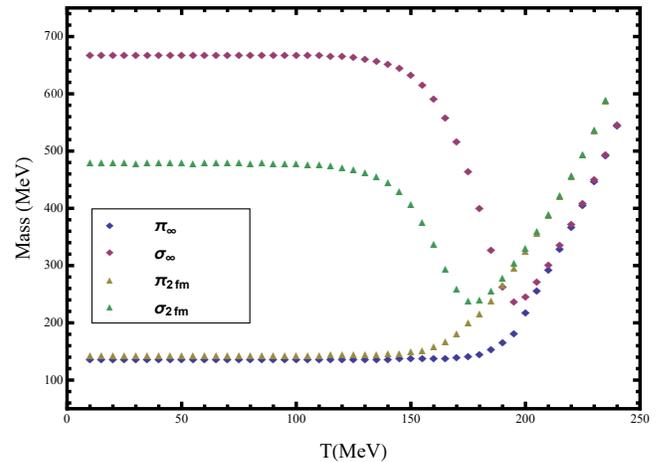}
\caption{The masses of $\pi$ and $\sigma$ as a function of $T$ at $\mu=0$ for there different volumes.}
\label{Fig:10}
\end{figure}

\begin{figure}
\includegraphics[width=0.47\textwidth]{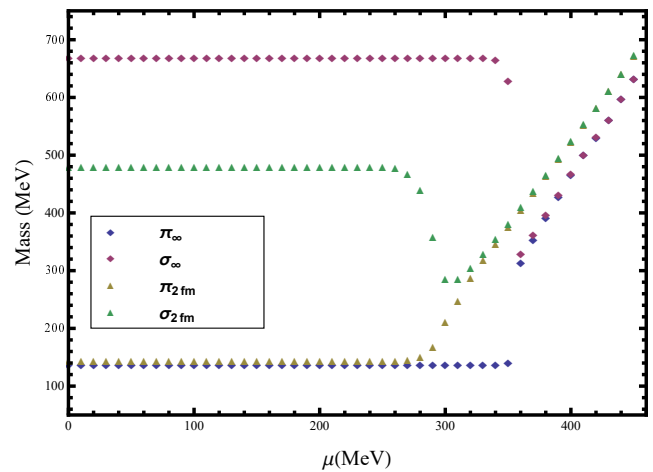}
\caption{The masses of $\pi$ and $\sigma$ as a function of $\mu$ at $T=0$ for there different volumes.}
\label{Fig:11}
\end{figure}

The $\pi$ and $\sigma$ are of particular interest, since they are directly associated with the chiral symmetry. As the lightest chiral partners, in the chiral symmetry broken phase they are quite different but they become degenerate when the chiral symmetry gets restored. From Fig. \ref{Fig:10} and Fig. \ref{Fig:11}, we can see clearly how it happens. We also notice that for low temperature and chemical potential with decreasing volumes the mass of $\pi$ will increase from $138\ \mathrm{MeV}$ to $145\ \mathrm{MeV}$, but the mass of $\sigma$ will decrease quite fast from $670\ \mathrm{MeV}$ to $480\ \mathrm{MeV}$. Actually, as the volume getting smaller and smaller, the masses of both of them getting closer and closer. This is another piece of evidence that the chiral symmetry breaking effects reduce with decreasing volumes. We note here that similar variation of $\pi$ and $\sigma$ with decreasing volumes has also been found in~\cite{PhysRevD.87.054009}.

\section{Summary and Conclusion}\label{summary}
Within the framework of $2+1$ PNJL model and by means of MRE we have studied chiral phase transition and masses of mesons inside a finite spherical volume. Our results show that, if the radius of spherical volume is larger than $10\ \mathrm{fm}$, finite volume effects are negligible.
 In the radius $2\ \mathrm{fm}\sim10\ \mathrm{fm}$, we find that the CEP shifts rapidly toward smaller temperatures but almost stay constant quark chemical potential, $\mu\simeq312\ \mathrm{MeV}$, when the radius decreases. This is an encouraging fact for the CEP search in heavy-ion collision experiments because of to obtain such high densities one needs to collide the ions at low $\sqrt{s}$, which means the temperature attained is lower. Especially, when the radius is smaller than $2\ \mathrm{fm}$, the whole phase diagram becomes a crossover which indicates that there is no CEP at all. Our shifting pattern of the CEP location in terms of volume agrees with other model calculations, i.e., the Quark-Meson model and Dyson-Schwinger equations, but a little difference is also observed.

About the finite volume effects on masses of mesons, we find for $\pi$ and $K$, their masses increase with decreasing volumes. But for $\sigma$, $\eta$ and $\eta '$ the situation is just the opposite. Especially, for the lightest chiral partners, $\pi$ and $\sigma$, in the chiral broken phase they are quite different but they become degenerate when chiral symmetry gets restored. Actually one can see a trend that the masses of this two chiral partners becoming closer to each other with decreasing volume, which means that the chiral symmetry breaking effects reduce with decreasing volumes. we also find, to some extent, that the decrease of volume restores the spontaneous breaking of chiral symmetry in a similar way as increase in temperature.

What we should pay attention to here is that our study is based on the mean-field approximation. This means we don't take effects of quark and meson fluctuations into account. Moreover, we only considered quark condensate in this work. If the diquark condensate is taken into account, it may affect the structure of the phase diagram and the location of the CEP further~\cite{PhysRevD.75.034007,PhysRevD.85.074007,STRODTHOFF2014350}. In summary, our present investigation already shows that the finite volume effects have considerable influences on the location of the CEP. It could be meaningful to lattice simulations as well as future experimental search for the CEP.
\acknowledgments
This work is supported by the National Natural Science Foundation of China (under Grants No. 11475085, No. 11535005, No. 11690030, and No. 11747140).

\bibliography{EPJC}
\end{document}